# Structural Distortions At Polar Manganite Interfaces

S. Koohfar[1], A. S. Disa[2], M. Marshall[2], F. J. Walker[2] C. H. Ahn[2] and D. P. Kumah[1*]

[1]Department of Physics, North Carolina State University

Raleigh, North Carolina, 207606, USA

[2]Center for Research on Interface Structures and Phenomena,

Yale University, New Haven, Connecticut 06520, USA

*Corresponding Author: dpkumah@ncsu.edu

**Abstract**

Electronic, lattice, and spin interactions at the interfaces between crystalline complex transition metal oxides can give rise to a wide range of functional electronic and magnetic phenomena not found in bulk. At hetero-interfaces, these interactions may be enhanced by combining oxides where the polarity changes at the interface. The physical structure between non-polar $SrTiO_3$ and polar $La_{1-x}Sr_xMnO_3(x=0.2)$ is investigated using high resolution synchrotron x-ray diffraction to directly determine the role of structure in compensating the polar discontinuity. At both the oxide-oxide interface and vacuum-oxide interfaces, the lattice is found to expand and rumple along the growth direction. The $SrTiO_3/La_{1-x}Sr_xMnO_3$ interface also exhibits intermixing of La and Sr over a few unit cells.

Coupling of the structural, electronic, spin and orbital degrees of freedom at the artificial interfaces formed between dissimilar complex-oxide perovskites layers has led to the emergence of a wide range of interesting physical properties including electronic and magnetic ordering and superconductivity, not found in the bulk constituent materials.[1] Of particular interest are the unique physical properties which have been discovered at interfaces between polar and non-polar complex transition metal oxides including the high-mobility two-dimensional electron gas formed at the polar $LaAlO_3$/non-polar $SrTiO_3$ interface.[2-4] At these interfaces where a polar discontinuity exists, lattice, chemical and/or electronic reconstructions can occur to alleviate the interfacial polar field and avoid the 'polar catastrophe'.[5-8] Understanding the relative contributions of these relaxation mechanisms at complex oxide interfaces and their relation to functional electronic and magnetic properties is essential for designing novel complex oxide heterostructures and manipulating emergent properties at these interfaces.

A model system in which these interactions can be linked to functional electronic and magnetic properties is the non-polar/polar interface between $SrTiO_3$(STO) and $La_xSr_{1-x}MnO_3$ ($0<x<1$) (LSMO). $A_{1-x}B_xMnO_3$ rare-earth manganite thin films are studied for their scientific and technological applications due to their half-metallic and ferromagnetic properties which can be effectively tuned by temperature, doping and strain. For a given doping, x, electronic and magnetic interactions depend structurally on the Mn-O-Mn bond angle and bond-lengths which control the Mn one-electron bandwidth.[9, 10] For thin LSMO films, a suppression has been observed in the magnetic and transport properties for films below a critical thickness of 4-10 unit cells (1 uc ~0.4 nm) when grown on nonpolar $SrTiO_3$ substrates. [8, 11, 12] The presence of an electronic and magnetic 'dead layer' at the film-substrate interface has been proposed to explain the thickness-dependent transition. Models proposed to explain the thickness-dependent transitions, including the interfacial/surface segregation of oxygen vacancies, interfacial chemical interdiffusion, charge transfer and unique strain driven interfacial orbital reconstructions.[8, 13-17]

Like many complex oxide materials with the $ABO_3$ perovskite crystal structure, LSMO films exhibit a polar stacking along the [001] growth direction and a polar discontinuity will exist at interfaces with non-polar substrates.[13, 18, 19] In similar polar systems , e.g. $LaNiO_3/LaAlO_3$[20] and $LaAlO_3/SrTiO_3$,[5, 7, 21] theoretical and experimental results indicate that the polar discontinuities that exist at the interfaces and surfaces in these systems may give rise to structural distortions involving polar displacements of the metal cations and oxygen anions to compensate the polar field. Recent theoretical reports indicate that similar ferrodistortive instabilities are present at the polar LSMO surface and the LSMO/STO interface, extending 2-4 unit cells away from the surfaces and interfaces.[19, 22] Imaging these polar structural distortions which couple to Mn-O bond parameters, will provide an understanding of the origin of the thickness dependent magnetic and electronic transitions in LSMO thin films. Additionally, by coupling external fields to these polar modes, multiferroic effects can be achieved. Hence, investigating possible structural reconstructions at these interfaces is crucial to engineering the electronic and magnetic properties of ultra-thin oxide films.

In this work, we use synchrotron diffraction and Coherent Bragg Rod Analysis (COBRA)[23-25] to obtain a detailed atomic scale structure of LSMO thin films grown epitaxially on $SrTiO_3$ (STO) substrates using oxygen plasma-assisted molecular beam epitaxy. We observe 3 main features that relieve the polar discontinuity at the STO/LSMO interface: a) polar cation-anion displacements along the growth direction at the LSMO-STO interface and LSMO/vacuum interface, b) a dilation of the *c* axis of the interfacial STO and LSMO layers and the surface LSMO layers and c) interdiffusion of La and Sr at the film/substrate interface. Here, by using a direct x-ray phase retrieval method combined with the analysis of superstructure rods, we determine structural distortions to the Mn-O bond related to thickness-dependent transport and magnetic transitions observed in this system.

A 10 uc $La_{0.8}Sr_{0.2}MnO_3$ film was grown by plasma assisted molecular beam epitaxy on a $TiO_2$-terminated STO substrate at a growth temperature of 750º C. After growth, the film was cooled down to room temperature in 1e-6 Torr oxygen plasma. The film thickness was monitored in-situ using reflection high energy electron diffraction (RHEED). The films were annealed *ex-situ* in flowing $O_2$ for 6 hours at 600ºC to minimize oxygen vacancies. Atomic force microscope images of the film show atomically smooth surface with unit cell step heights (~4 A).

To determine the atomic scale structure of the films, surface x-ray diffraction measurements were carried out at the 33ID beamline at the Advanced Photon Source. Crystal truncation rods (CTRs) and half-order superstructure rods were measured for the samples along rod crystallographic directions defined by the bulk $SrTiO_3$ substrate lattice. CTRs were measured with an incident x-ray energy of 16 keV. The x-ray phases associated with the CTR data were determined using the coherent Bragg rod analysis technique (COBRA) to obtain the three-dimensional atomic structures of the STO/ LSMO heterostructure. Figure 1(a) and 1(b) shows representative measured CTRs and superstructure rods, respectively, and associated fits for the nominal 10 uc LSMO film. The fits are obtained from a structural refinement of the COBRA determined structures using the genetic-algorithm based GenX program[26]. From the refined structures, we obtain for each atomic layer of the film and interfacial layers of the substrate, the ionic composition profile and positions with picometer-scale resolution. The relevant features to be discussed are the layer spacings, the compositional profile, cation-anion vertical displacements and layer-resolved oxygen octahedral rotations.

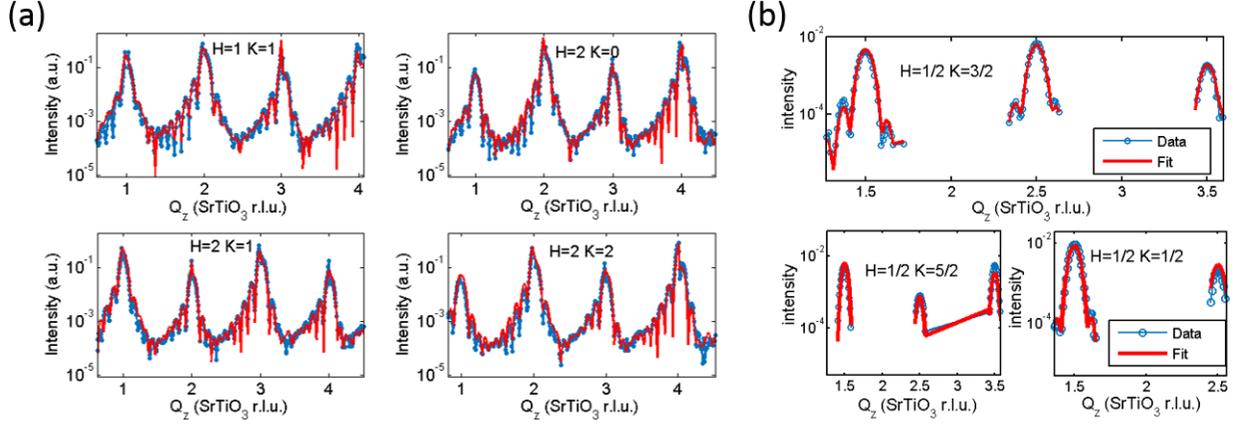

*Figure 1: (a) Crystal truncation rods (CTRs) for a nominal 10 unit cell La$_{0.8}$Sr$_{0.2}$MnO$_3$ film grown on (001)-oriented SrTiO$_3$. Measured data (blue circles) and best fit (red) are shown. (b) Half order superstructure rods for a nominal 10uc La$_{0.8}$Sr$_{0.2}$MnO$_3$ film grown on SrTiO$_3$. The layer-resolved octahedral rotations are determined by fits (red lines) to the measured diffraction intensities (blue circles).*

Figure 2(a) shows the vertical spacing as a function of distance from the nominal film/substrate interface between adjacent A-site (La/Sr) layers for the LSMO film and the top layers of the substrate determined from the fits in Figure 1. Bulk La$_{0.8}$Sr$_{0.2}$MnO$_3$ (pseudocubic bulk c= 3.89 Å) and is closely lattice matched with the STO (c=3.905 Å) substrate. The film is expected to be under slight tensile strain ($e = \frac{c_{LSMO} - c_{STO}}{c_{STO}} = -0.4\%$) on STO . We observe an unexpected dilation of the out-of-plane c lattice constant for the surface LSMO layer and within the STO and LSMO layers at the film/substrate interface. The inner LSMO layers (10 Å<z<30 Å) have expected contractions in the lattice spacing to ~3.88 Å due to the 0.4% tensile strain exerted by the STO substrate. This interfacial dilation has previously been observed and interpreted as arising due to a suppression of octahedral rotations at the STO/LSMO interface[27] and/or a consequence of La-Sr intermixing leading to an increase in trivalent Ti and Mn at the interface.[16]

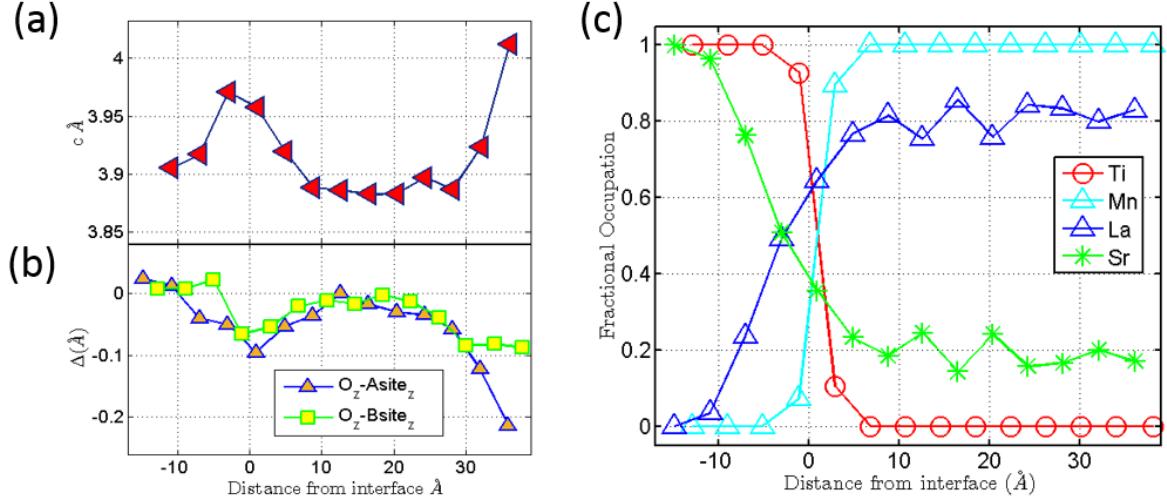

*Figure 2. Structural results for a nominal 10uc $La_{0.8}Sr_{0.2}MnO_3/SrTiO_3$ interface as a function of layer distance from the nominal film/substrate interface. The nominal film-substrate interface is at z=0 Å. (a) Profile of out-of-plane lattice parameter, c, determined from the vertical distance between A-site(La/Sr) ions and (b)Vertical displacements of oxygen ions relative to cations(A-site(La/Sr), B-site(Ti/Mn) ) in the same plane. (c) Fractional ionic composition as a function of layer distance from the nominal interface.*

Along the [001] direction in bulk LSMO, the net displacements of the cations ($La^{3+}$, $Sr^{2+}$ and $Mn^{3+}$-$Mn^{4+}$) and $O^{2-}$ anions in the same planes are zero. Figure. 2(b) shows the measured anion-cation displacements measured in the AO and $BO_2$ layers along the [001] growth direction in LSMO/STO. Deep in the substrate ( z<-10 Å), the displacements are 0 as expected for bulk STO. The anion relative displacements are found to be negative (i.e. $O^{2-}$ ions move towards the substrate relative to the respective cations) for the 2 top STO layers and the next 3 film layers adjacent to the interface. This polarization coincides with the regions of the film where the unit cell contracts along the c-axis. The LSMO layers between 10 Å and 25 Å have no polar displacements, and a gradual increase in the amplitude of the distortions is observed in the top 4 film surface layers with the oxygen ions displaced relative to the cation vertical positions towards the film/substrate interface.

We observe from the COBRA-derived electron density profiles that the interface between the STO substrate and the LSMO film is not abrupt. By fitting the ionic fractional occupations at the interfaces, a chemical profile is obtained and shown in Figure 2(c). La interdiffusion occurs into the top 3 substrate unit cells leading to the substitution of $Sr^{2+}$ with $La^{3+}$ in the STO and $Sr^{2+}$ enrichment in the interfacial LSMO layer. No significant Sr surface segregation is observed, in contrast to previous reports of Sr surface segregation in LSMO films.[16, 28, 29] Sr-rich surfaces have been found to be associated with surface oxygen vacancies, hence, we conclude that oxygen vacancies are minimized in our films.

The lattice polarization ($\Delta$) profile is consistent with a theoretically predicted surface ferrodistortive instability for $MnO_2$-terminated LSMO [22] and ionic rumpling at LSMO/STO interfaces[19] driven by the interfacial and surface polar discontinuities. We note that the direction, amplitude and decay length of the polar distortions in the LSMO and the extension into the STO substrate are in agreement with theoretical predictions.[19] Additionally, we find that the interfacial polar distortions coincide with the expansion of the unit lattice in the vertical direction in both the STO substrate and the interfacial LSMO region.

To understand the observed structural and chemical profiles, we consider a simple model involving the net charges of the film and substrate along the [001] growth direction. The $SrTiO_3$ substrate consists of neutral alternating $[Sr^{2+}O^{2-}]$ and $[Ti^{4+}O^{4-}]$ planes. For the LSMO films, the Mn site has a mixed valance state of $x$ $Mn^{4+}$ and $(1-x)$ $Mn^{3+}$ so, the net charges on the $[(1-x)La^{3+}xSr^{2+}O^{2-}]$ and $[Mn^{3+/4+}O^{2-}]$ are $+0.8e$ and $-0.8e$ respectively for $x=0.2$. A polar discontinuity will exist at the interface between the neutrally charged planes of the STO substrate and the LSMO film and at the LSMO/vacuum interface. At the film-substrate interface, intermixing between $Sr^{2+}$ and $La^{3+}$ leads to a gradual transition from the neutral planes of the STO substrate and the charged planes in the LSMO film reducing the polar discontinuity. [16, 30] The presence of structural polar distortions indicates that the chemical diffusion is not sufficient to completely compensate the interfacial polar field. Indeed, attempts to control the interfacial intermixing driven by the polar discontinuity by modifying the composition of the initial LSMO layer, show that the interfacial intermixing can be reduced leading to improved electronic and magnetic properties in ultrathin manganite films.[13, 30, 31]

Due to the strong coupling of the structural and electronic and magnetic properties of the manganites, the observed distortions are expected to play a role in the reported 'dead-layer' effects observed in thin manganite films in addition to the interfacial chemical intermixing. Changes to the Mn-O bonding arising from the polar distortions (which effectively increase the Mn-O bond length and may couple to O-octahedral rotations) are expected to induce changes in the one-electron bandwidth, $W$, which controls the super-exchange process of spin and charge exchange in the system where $W = \cos(\frac{180-\theta_{B-O-B}}{2})/d_{B-O}^{3.5}$.[10, 32, 33] The combination of polar distortions and octahedral rotations leads to 2 distinct values for the bond angles and bond lengths along a B-O in-plane chain as shown in the inset of Figure 3(a). The measured averaged layer-resolved in-plane B-O-B angles (B=Ti,Mn), $\theta_{B-O-B}$, and B-O bond lengths, $d_{B-O}$, determined from fits to the superstructure rods in Figure 1(b) are plotted in Figures 3(a) and 3(b) respectively. The bond angles decrease gradually from 180° in the STO substrate to ~161° in the LSMO layers next to the interface followed by an increase to the bulk value of 164° and a decrease in the 2 surface LSMO layers. An increase in the B-O bond distance is also observed at the film interface and surface. The average bandwidth for each layer are plotted in Figure 3(c). The 3 interfacial LSMO layers and top 2 layers with W< bulk are expected to have reduced conductivity relative to bulk.

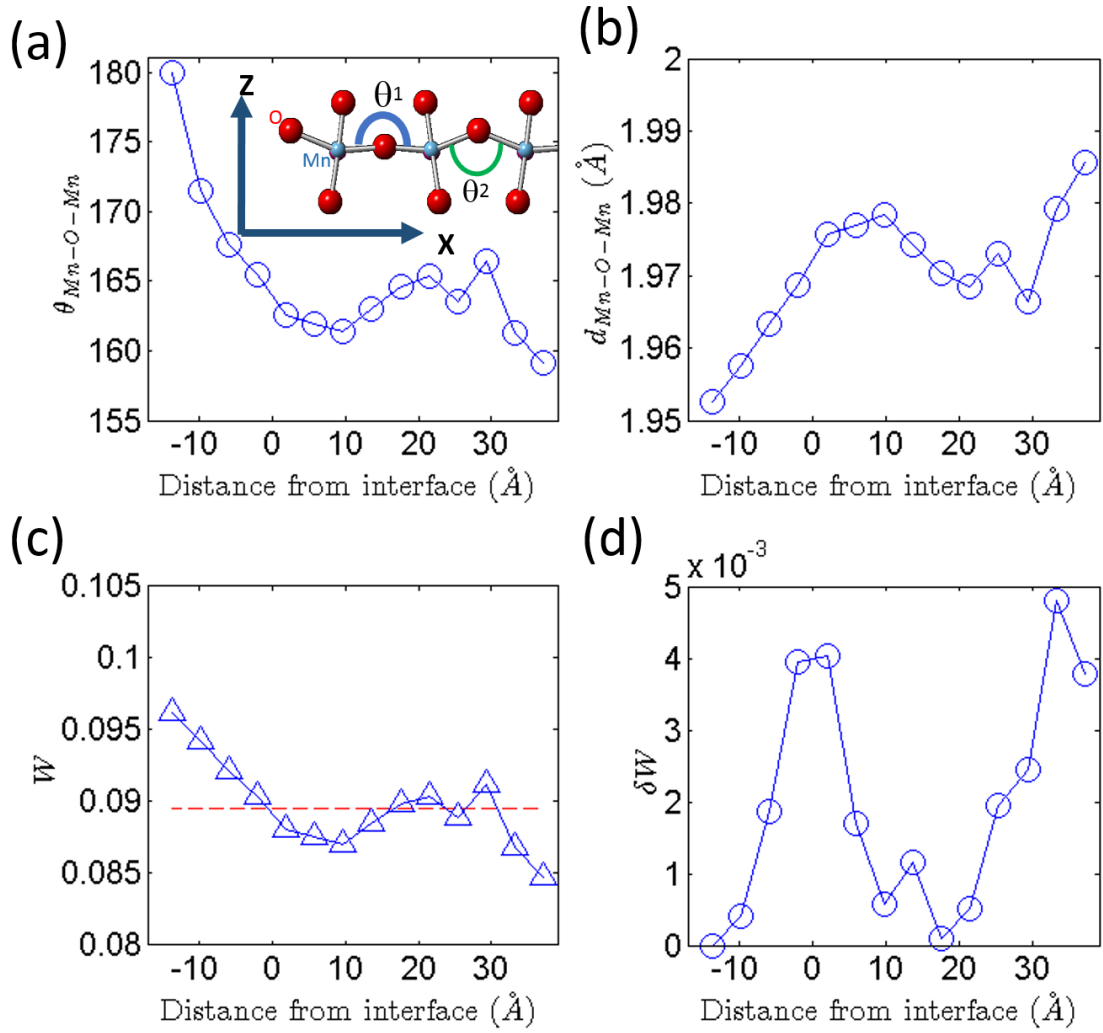

Figure 3. Layer resolved in-plane B-O-B (B=Ti,Mn) bond parameters. (a) In-plane averaged B-O-B bond angles $\theta_{Mn-O-Mn}$. The inset illustrates the effect of the polar displacements on the B-O-B bond angles in neighboring unit cells along the x-direction. (b) In-plane averaged bond distances, $d_{Mn-O-Mn}$ along a B-O-B chains along (c) The resulting one-electron bandwidth, $W = cos(\frac{180-\theta_{B-O-B}}{2})/d_{B-O-B}^{3.5}$. The dashed line indicates W for bulk $La_{0.8}Sr_{0.2}MnO_3$ (d) Bandwidth difference $\delta W$, for neighboring O ions along the [010] direction.

As mentioned above, the polar distortions coupled to the octahedral tilts leads to slightly different *W*'s along the in-plane axes. We calculate the differences in bandwidth, $\delta W$, for each layer in Figure 3(d). As expected, an increase in $\delta W$ exists at the film/substrate interface and the film surface, where large polar distortions are present. A consequence of $\delta W > 0$, is that neighboring O anions in that layer will have different bonding environments. Additional theoretical calculations and experimental measurements will be needed to determine if this leads to charge disproportionation on the anion sites and the effect on superexchange interactions in complex oxides and/or the opening of an electronic gap similar to Jahn-Teller distortions.

In conclusion we have shown that a combination of structural and chemical interactions occur at the polar manganite interfaces to compensate the polar discontinuity. Polar distortions and chemical intermixing are observed in thin LSMO films grown on non-polar STO substrates using synchrotron x-ray diffraction CTR measurements. The observed changes in structure and composition arise due to interfacial and surface polar discontinuities and are linked to a suppression of the electronic and magnetic properties of thin LSMO films. These results have important implications for the design of multiferroic oxide thin films where the observed interfacial polar instability can be engineered to modulate magnetism and electronic transport.

Use of the Advanced Photon Source was supported by the U. S. Department of Energy, Office of Science, Office of Basic Energy Sciences, under Contract No. DE-AC02-06CH11357. Yale acknowledges support from FAME and ONR.